\documentclass[a4paper,11pt]{article}
\usepackage{jheppub} 
\usepackage{lineno}
\date{}

\title{Global Type IIA Vacuum: A Unified Model for the Standard Model and Cosmology}

\author[1,2,3]{Yang Liu}
\affiliation[1]{Department of Physics, Tsinghua University, Beijing 100084, China}
\affiliation[2]{School of Physics and Astronomy, University of Nottingham, University Park, Nottingham NG7 2RD, United Kingdom}
\affiliation[3]{Nottingham Centre of Gravity, University of Nottingham, Nottingham NG7 2RD, UK}

\emailAdd{liu-yang\_1990@mail.tsinghua.edu.cn}

\abstract{We propose a unified framework within Type IIA string theory, based on a globally consistent intersecting D6-brane model compactified on a $T^6/(\mathbb{Z}_2 \times \mathbb{Z}_2)$ orientifold. The model realizes the MSSM-like spectrum providing a framework for addressing four fundamental problems: CP violation originates from both geometric phases in Yukawa couplings and non-perturbative phases induced by E2-instantons; the observed baryon asymmetry arises via instanton-mediated 
operators combined with moduli-driven leptogenesis; the electroweak hierarchy is stabilized through controlled SUSY breaking with a TeV-scale gravitino mass near a metastable vacuum; and a de Sitter uplift is achieved via anti-D6-branes in an STU moduli stabilization scheme. Crucially, the interplay of intersecting brane geometry, Euclidean D2-instantons, and flux-induced moduli potentials provides a coherent mechanism linking collider, flavor, and cosmological phenomena.}

\begin{document}
\date{}
\maketitle
\flushbottom

\section{Introduction}
\label{sec:intro}

Understanding the origin of fundamental asymmetries and scales in particle physics and cosmology remains one of the central challenges of high-energy theory. Despite the tremendous success of the Standard Model (SM), it fails to account for several observed phenomena, such as the source and structure of CP violation, the matter–antimatter asymmetry of the Universe (quantified by $\eta_B \sim 6 \times 10^{-10}$), the hierarchy between the electroweak and Planck scales, and the presence of a small but positive cosmological constant associated with de Sitter (dS) space. These puzzles are not isolated: their resolutions likely stem from a deeper and more unified framework that extends beyond the SM.

In the SM, CP violation arises from a single physical phase in the CKM matrix. However, this mechanism is insufficient to generate the observed baryon asymmetry via electroweak baryogenesis. Furthermore, the SM lacks any explanation for neutrino masses, leptonic CP phases, or mechanisms capable of accommodating the observed dark energy scale. These deficiencies strongly suggest that a UV-complete theory with extended symmetry, new degrees of freedom, and higher-dimensional structure is needed to coherently address these questions.

String theory provides a natural arena to explore such unified explanations. It offers not only a consistent framework for quantum gravity but also rich geometric and topological structures -- such as intersecting D-branes, background fluxes, and stringy instantons -- that can dynamically generate hierarchical scales, break global symmetries, and induce effective operators beyond the renormalizable SM. Among the most promising constructions are Type IIA intersecting D6-brane models, which can yield MSSM-like spectra and simultaneously allow for moduli stabilization and axionic dynamics within the same compactification.

However, most prior studies focus on isolated aspects of this broader picture: CP violation in Yukawa couplings \cite{blumenhagen2008nonperturbative}, instanton-induced Majorana masses \cite{ibanez2007neutrino}, or the realization of dS vacua through uplifting mechanisms such as KKLT \cite{kachru2003sitter} or KL \cite{kallosh2004landscape}. Very few works attempt to integrate these components into a single, globally consistent string model that simultaneously explains multiple core phenomena. The lack of structural integration has limited progress toward a genuinely unified string phenomenology.

In our previous paper \cite{liu2025fine}, we proposed a unified string-theoretic framework in which three major fine-tuning problems -- the strong CP problem, the hierarchy problem, and the cosmological constant problem -- can be simultaneously addressed. We worked within Type IIA string theory compactified on a $T^6/(\mathbb{Z}_2 \times \mathbb{Z}_2)$ orientifold, incorporating RR and NS fluxes, and adopted a Kallosh-Linde-type structure for moduli stabilization. A natural question is that how to get a realistic Type IIA string theory model to describe our real universe?

In \cite{dvali2022strong}, the author pointed out that the prevailing understanding of quantum gravity is mainly based on its S-matrix formulation, which imposes strict constraints on the vacuum landscape of the theory. This framework specifically excludes the possibility of de Sitter vacua, i.e., vacua with positive energy densities. However, other cosmological spacetimes that do not asymptotically approach the Minkowski vacuum also pose challenges from the S-matrix perspective \cite{dvali2022strong}. In addition, in \cite{Liu:2023vqp,Liu:2024blx}, we demonstrated that Minkowski vacua are the most stable for a wide class of effective field theories. In fact, the cosmological constant of the Universe is extremely small, which means that the spacetime is very close to Minkowski spacetime \cite{copeland2006dynamics}. Therefore, in this paper, we use the fact (extremely small cosmological constant) and start with Minkowski spacetime, treating de Sitter and anti-de Sitter vacua as small perturbations around it.

We present a coherent framework based on a globally consistent Type IIA intersecting D6-brane model \cite{marchesano2004building,camara2005fluxes,ibanez2012string}, which we revisit and extend in this paper. This model not only realizes MSSM-like spectrum without gauge anomalies, but also permits the inclusion of E2-instanton effects, axion couplings, and moduli dynamics, enabling a unified explanation of:
\begin{itemize}
 \item CP violation via both CKM and non-CKM sources (including geometric phases from complex moduli and instanton-induced phases);
 \item Baryon and lepton number violation via stringy instantons generating $\Delta B = 1$ and $\Delta L = 2$ operators;
 \item The observed baryon asymmetry via moduli-driven leptogenesis in a time-dependent background;
 \item The electroweak hierarchy via controlled SUSY breaking and gravitino mass generation near a metastable Minkowski vacuum;
 \item The emergence of a dS vacuum through KL-type uplift with backreaction on moduli stabilization.
\end{itemize}

The core structure of this mechanism -- interweaving brane intersections, Euclidean D-brane instantons, axionic moduli, and background fluxes -- provides a blueprint for a phenomenologically rich and internally consistent string vacuum capable of addressing multiple fundamental problems simultaneously.

The construction presented here naturally gives rise to a defined research programme. Subsequent publications will delve into the detailed phenomenology of inflation, dark matter candidates, dark energy and black hole within this framework, the foundations for which are laid out in this work.

This paper is organized as follows. In Section 2, we review four-forms in Type IIA orientifolds and the D6-brane configuration of the model in this paper (we call it Model A), including its chiral spectrum, gauge structure, and consistency. Section 3 presents the derivation of CP-violating Yukawa structures, the geometric origin of complex phases, and the resulting CKM and Jarlskog parameters. In Section 4, we analyze E2-instanton configurations that induce $\Delta B$ and $\Delta L$ operators, compute their associated CP phases, and derive the effective $4D$ operators. Section 5 constructs a dynamical moduli-driven leptogenesis scenario, deriving the baryon asymmetry parameter $\eta_B$. In Section 6, we briefly review STU model in Type IIA string theory. In Section 7, we discuss moduli stabilization and compatibility with dS vacuum. In Section 8 we show an illustrative example.  In Section 9 we summarize the main results of this paper and outlook the open questions and future directions.

\section{Four-forms in Type IIA orientifolds, model setup and consistency conditions}
\label{sec:fIIAoMsacc}

In Section 2, we will first provide a brief review of the fundamentals of 4-forms in Type IIA orientifolds, followed by an introduction to Model A and a check of its consistency conditions.

\subsection{Review of 4-forms in Type IIA orientifolds}
We review the appearance of $4D$ 4-forms in Type IIA orientifold compactifications. The compactification of ten-dimensional massive Type IIA string theory on a Calabi-Yau threefold with background fluxes has been extensively examined in \cite{grimm2005effective, louis2002type, villadoro2005N, dewolfe2005type, camara2005fluxes}. In \cite{bielleman2015minkowski}, the authors carried out the same compactification, carefully tracking all Minkowski 4-forms that arise from dimensionally reducing the $10D$ RR and NSNS fluxes. This leads to a new formulation of the scalar potential in terms of Minkowski 4-forms \cite{bielleman2015minkowski}.

Our focus is on the role of Minkowski 3-form fields in the flux-induced scalar potential. In addition to the universal RR 3-form $C_3$, 3-forms can also be obtained by dimensionally reducing higher RR and NSNS fields, such as $C_5$, $C_7$, $C_9$, and $H_7$, while considering three of the indices in Minkowski space. We will adopt the democratic formulation \cite{bergshoeff2001new}, where all $p$-form fields $C_p$ with $p = 1, 3, 5, 7$ are included. Consequently, we will need to impose the Hodge duality relations:
\begin{equation} \label{Hodgeduality}
  G_6 = - \star_{10} G_4, \quad G_8 =  \star_{10} G_2, \quad G_{10} =  -\star_{10} G_0.
\end{equation}
At the level of the equations of motion, we impose conditions to prevent overcounting the physical degrees of freedom. As a result, we obtain $2 h^{(1,1)}_{-}+2$ Minkowski 4-forms: $F^0_4$, $F^i_4$, $F^a_4$, and $F^m_4$. Specifically, there are $h^{(1,1)}_{-}$ $F^i_4$ fluxes, $h^{(1,1)}_{-}$ $F^a_4$ fluxes, one $F^0_4$ flux, and one $F^m_4$ flux. Although the details are not covered in this paper, further information can be found in \cite{bielleman2015minkowski}. Additionally, the fields $B_2$ and $C_3$ can be expanded as follows:
\begin{equation} \label{B2C3basis}
  B_2 = \sum_i b_i \omega_i, \quad C_3= \sum_I c^I_3 \alpha_I.
\end{equation}
Here, $b_i$ and $c^I_3$ are 4D scalars that correspond to the axionic components of the complex supergravity fields $T$, $S$, and $U$, as expressed by the following:
\begin{equation} \label{ImTi}
 \text{Im} T_i= - \int B_2 \wedge \tilde{\omega}^i = -b^i, \quad i= 1,..., h^{(1,1)}_{-}
\end{equation}
\begin{equation} \label{ImUi}
 \text{Im} U_i= \int C_3 \wedge \beta^i = c^i_3, \quad i= 1,..., h^{3}_{+}
\end{equation}
\begin{equation} \label{ImS}
 \text{Im} S= -\int C_3 \wedge \beta^0 = -c^0_3,
\end{equation}
where $\tilde{\omega}^i$, $\beta^i$ and $\beta^0$ are the elements of the cohomology basis \cite{bielleman2015minkowski}.

It is known that, in addition to the standard RR and NS fluxes, there may also be other, less-explored NS fluxes. Among these are the geometric fluxes in toroidal models, which emerge in the context of Scherk-Schwarz reductions \cite{bielleman2015minkowski}. Geometric fluxes can be defined on a factorized 6-torus $T^6$, where O6-planes wrap 3-cycles. Furthermore, we assume a $\mathbb{Z}_2 \times \mathbb{Z}_2$ orbifold twist, which leads to the survival of only diagonal moduli after the projection. In this scenario, we are left with 3 Kähler moduli and 4 complex structure moduli (including the complex dilaton). This configuration involves 12 geometric fluxes $\omega^M_{NK}$, which can be efficiently organized into a 3-vector $a_i$ and a $3 \times 3$ matrix $b_{ij}$ \cite{bielleman2015minkowski}.

\subsection{Setup and consistency conditions of Model A}
In the Type IIA $T^6/(\mathbb{Z}_2 \times \mathbb{Z}_2)$ orientifold theory, the complete perturbative superpotential is given by
\begin{eqnarray} \label{Wf}
\begin{split} 
W =  & e_0 + ih_0 S + \sum^3_{i=1} [(ie_i -a_i S -b_{ii} U_i -\sum_{j\neq i} U_j)T_i- ih_i U_i]  \\
& -q_1 T_2 T_3 - q_2 T_1 T_3 - q_3 T_1 T_2 + im T_1 T_2 T_3,
\end{split}
\end{eqnarray}
where $e_0$, $e_i$, $h_0$, $h_i$, $a_i$, $b_{ij}$, $q_i$, and $m$ are flux parameters \cite{camara2005fluxes}. The Kähler potential has the standard form
\begin{equation} \label{Kp}
 K= - \ln (S + S^{*}) - \sum^{3}_{i=1} \ln (U_i + U^{*}_i) - \sum^{3}_{i=1} \ln (T_i + T^{*}_i). 
\end{equation}
Here, $S$ denotes the axio-dilaton, $U_i$ represents the complex structure moduli, and $T_i$ corresponds to the volume (Kähler) moduli \cite{camara2005fluxes}.

In this paper, we focus on a specific case of the Type IIA $T^6/(\mathbb{Z}_2 \times \mathbb{Z}_2)$ orientifold theory: a 3-generation $\mathcal{N}=1$ MSSM-like model \cite{camara2005fluxes}, which we refer to as Model A. The corresponding perturbative superpotential is given by
\begin{equation} \label{SW}
 W = -T_2 (a_2 S + b_{21}U_1) - T_3 (a_3 S + b_{31}U_1) + e_0 + i h_0 S -ih_1 U_1 + ie_2 T_2 + ie_3 T_3, 
\end{equation}
where $h_0$ and $h_1$ are NSNS fluxes, and $e_0$, $e_2$, and $e_3$ are RR fluxes \cite{camara2005fluxes}. The fluxes $q_i$ and $m$ are set to zero.

In general, we consider stacks of $N_a$ intersecting D6-branes wrapping the factorizable 3-cycle
\begin{equation} \label{Pia}
\Pi_a = (n^1_a, m^1_a) \otimes (n^2_a, m^2_a) \otimes (n^3_a, m^3_a),
\end{equation}
along with their corresponding orientifold images, which wrap the cycles $\otimes_i (n^i_a, -m^i_a)$. Here, $n^i_a$ and $m^i_a$ denote the wrapping numbers along the $x^i$ and $y^i$ directions of the $i$-th two-torus, respectively. In the case of the $\mathbb{Z}_2 \times \mathbb{Z}_2$ IIA orientifold, the RR tadpole cancellation conditions in the presence of fluxes take the following form \cite{camara2005fluxes}:
\begin{equation} \label{tcc11}
\sum_a N_a n^1_a n^2_a n^3_a + \frac{1}{2} (h_0 m + a_1 q_1 + a_2 q_2 + a_3 q_3) =16,
\end{equation}
\begin{equation} \label{tcc21}
\sum_a N_a n^1_a m^2_a m^3_a + \frac{1}{2} (m h_1 - q_1 b_{11} - q_2 b_{21} - q_3 b_{31}) =-16,
\end{equation}
\begin{equation} \label{tcc31}
\sum_a N_a m^1_a n^2_a m^3_a + \frac{1}{2} (m h_2 - q_1 b_{12} - q_2 b_{22} - q_3 b_{32}) =-16,
\end{equation}
\begin{equation} \label{tcc41}
\sum_a N_a m^1_a m^2_a n^3_a + \frac{1}{2} (m h_3 - q_1 b_{13} - q_2 b_{23} - q_3 b_{33}) =-16.
\end{equation}
For the case where $q_i = m = 0$, the RR tadpole cancellation conditions for Model A become
\begin{equation} \label{tcc12}
\sum_a N_a n^1_a n^2_a n^3_a  =16,
\end{equation}
\begin{equation} \label{tcc22}
\sum_a N_a n^1_a m^2_a m^3_a  =-16,
\end{equation}
\begin{equation} \label{tcc32}
\sum_a N_a m^1_a n^2_a m^3_a  =-16,
\end{equation}
\begin{equation} \label{tcc42}
\sum_a N_a m^1_a m^2_a n^3_a  =-16.
\end{equation}
The value $(-16)$ in the last three conditions corresponds to the RR tadpole contribution from the three remaining orientifold planes present in the $\mathbb{Z}_2 \times \mathbb{Z}_2$ setup \cite{camara2005fluxes}.

\begin{table}[ht]
\centering
\caption{Wrapping numbers giving rise to a MSSM-like spectrum. Branes $h_1$, $h_2$ and $o$ are added in order to cancel RR tadpoles.} 
\begin{tabular}{|c|c|c|c|}
\hline
$ N_{i} $ & $ (n_{i}^{1},m_{i}^{1}) $ & $ (n_{i}^{2},m_{i}^{2}) $ & $ (n_{i}^{3},m_{i}^{3}) $ \\
\hline
$ N_{a}=8 $ & (1,0) & (3,1) & (3,-1) \\
\hline
$ N_{b}=2 $ & (0,1) & (1,0) & (0,-1) \\
\hline
$ N_{c}=2 $ & (0,1) & (0,-1) & (1,0) \\
\hline
$ N_{h_{1}}=2 $ & (-2,1) & (-3,1) & (-4,1) \\
\hline
$ N_{h_{2}}=2 $ & (-2,1) & (-4,1) & (-3,1) \\
\hline
$ 8N_{f} $ & (1,0) & (1,0) & (1,0) \\
\hline
\end{tabular}
\end{table}
Table 1 presents the setup and wrapping numbers of Model A. The branes $a$, $b$, and $c$ give rise to a 3-generation MSSM-like spectrum, while the additional branes $h_{1,2}$, as shown in Table 1, are employed to help cancel the RR tadpoles. It is important to note that, since $m = q_i = 0$, the fluxes do not contribute to the RR tadpole in this background. Therefore, one can consider the addition of D6-branes as in the case with $N_f = 5$ in Table 1 \cite{camara2005fluxes}.

For Model A, it can be verified that the branes $a$, $b$, and $c$, where the Standard Model is located, trivially satisfy the Freed-Witten (FW) constraint. However, the branes of type $h_{1,2}$ may pose issues unless the following condition is met:
\begin{equation} \label{FWh12}
a_2(m^1_a m^2_a m^3_a)- b_{21}(m^1_a n^2_a n^3_a)=a_2 -12b_{21}=0,
\end{equation}
which can be easily satisfied by appropriately choosing $a_2$ and $b_{21}$ \cite{camara2005fluxes}. Further discussions on the Freed-Witten anomaly can be found in \cite{freed1999anomalies}.

Model A, based on intersecting D6-branes on a toroidal orientifold, meets several key consistency conditions required for a globally consistent MSSM-like construction:\\
\textbf{1. Anomaly cancellation}\\
\begin{itemize}
\item RR tadpole cancellation is achieved by an appropriate choice of visible and hidden D6-branes wrapping factorizable 3-cycles \cite{camara2005fluxes}. For example, according to Table 1, we have 
\begin{eqnarray} \label{RRcheck}
\begin{split} 
\sum_a N_a n^1_a n^2_a n^3_a =  & 8 \times 1 \times 3 \times 3 + 2 \times 0 \times 1 \times 0 + 2 \times 0 \times 0 \times 1  \\
& + 2 \times (-2) \times (-3) \times (-4) + 2 \times (-2) \times (-4) \times (-3) \\
& + 8 \times 5 \times 1 \times 1 \times 1 =16,
\end{split}
\end{eqnarray}
namely, \eqref{tcc12}. Similarly, we can check that Model A satisfies \eqref{tcc22}-\eqref{tcc42} as well.
\item 
The chiral spectrum arising from the intersecting brane configuration of Model A naturally realizes a left-right symmetric extension of the MSSM, with gauge group $SU(3)_C \times SU(2)_L \times SU(2)_R \times U(1)_{B-L} \times [U(1)]$. This structure possesses a built-in mechanism for anomaly cancellation \cite{mohapatra1975left,pati1974lepton}. The non-Abelian anomalies for $SU(3)_C$, $SU(2)_L$, and $SU(2)_R$ vanish due to the chiral-safe property of the representations and the fact that the number of doublets is even. The cubic and mixed anomalies involving the Abelian $U(1)$ factors are canceled via the generalized Green-Schwarz mechanism \cite{ibanez2012string}, which is intrinsic to string theory.
\item 
The anomalous $U(1)$ gauge bosons acquire Stückelberg masses by absorbing axionic modes from the RR sector, decoupling from the low-energy theory.
\item The hidden sector is introduced to cancel RR tadpoles but does not intersect chirally with the visible sector, hence avoiding exotic matter \cite{blumenhagen2007four,cvetivc2001three,ibanez2001getting}.
\end{itemize}
\textbf{2. K-Theory constraints}\\
Tadpole cancellation is a necessary but not always sufficient condition for global consistency. $Z_2$-valued K-theory constraints are a more subtle requirement that detects stable stringy solitons that would otherwise lead to global anomalies \cite{ibanez2012string,uranga2003chiral,marchesano2007progress}. It can be easily checked that Model A satisfies the following discrete K-theory constraints:
\begin{equation} \label{dKtc}
\sum_a N_a m^1_a m^2_a m^3_a \in 4 \mathbf{Z}, \quad \sum_a N_a n^1_a n^2_a m^3_a \in 4 \mathbf{Z}, \quad \text{and permutations.}
\end{equation}
For example, according to Table 1, we have 
\begin{eqnarray} \label{Kcheck}
\begin{split} 
\sum_a N_a m^1_a m^2_a m^3_a =  & 8 \times 0 \times 1 \times (-1) + 2 \times 1 \times 0 \times (-1) + 2 \times 1 \times (-1) \times 0  \\
& + 2 \times 1 \times 1 \times 1 + 2 \times 1 \times 1 \times 1 + 8 \times 5 \times 0 \times 0 \times 0 =4\\
& \in 4 \mathbf{Z}.
\end{split}
\end{eqnarray}
Similarly, we can check that Model A satisfies other K-theory constraints.\\
\textbf{3. Supersymmetry conditions}\\
In order to preserve $\mathcal{N}=1$ supersymmetry, we should require the SUSY condition:
\begin{equation} \label{SUSYcon}
\theta_1 + \theta_2 + \theta_3 = 0 \quad \text{mod} \quad 2\pi,
\end{equation}
where $\theta_i = \tan^{-1}(\frac{m^i R_2}{n^i R_1})$ and $R_1$, $R_2$ are the two radii along two directions of every $T^2_i$. In principle, we can always choose the parameter $U_i=R^{(i)}_2/R^{(i)}_1$ to satisfy the SUSY condition \eqref{SUSYcon}. 
To conclude, the Model A is a global embedding model.

\subsection{The role of a small flux parameter: $m \neq 0$}
The tadpole cancellation conditions provide a stringent set of constraints on any consistent string compactification. In our framework, achieving MSSM-like spectrum often requires specific brane configurations that contribute a significant excess to the RR tadpoles. This excess must be precisely canceled by the introduction of hidden sector branes, as indicated in our Model A (see Table 1). We briefly discuss the role of a small flux parameter $m \neq 0$.

\begin{itemize}
\item \textbf{The $m=0$ case:} Setting the NSNS flux parameter $m = 0$ represents a specific, highly symmetric point in the discretum of possible flux vacua \cite{douglas2003statistics}. In this case, the entire burden of tadpole cancellation falls upon the precise winding numbers of the visible and hidden D6-brane stacks. This can place severe restrictions on model building.
\item \textbf{The $m \neq 0$ case:} Allowing for a small, non-zero integer value for $m$ acts as a controlled perturbation that provides a crucial degree of flexibility \cite{marchesano2007progress}. It does not alter the topological structure of the compactification, the chiral spectrum, or the gauge group. 
We will discuss more details in Section 7.
\end{itemize}
Therefore, in our analysis, we consider the case of a small, non-zero $m$ as the generic situation. It represents a mild perturbation that leaves the core structure of Model A -- its gauge group, chiral matter content, and unified mechanism for solving the hierarchy, strong CP, and cosmological constant problems -- completely intact. Its primary role is to enable the precise fine-tuning of the vacuum energy and the supersymmetry breaking scale, which are the final, critical steps in realizing a de Sitter vacuum compatible with all observations.

\section{CP violation from Yukawa and instanton phases}
\label{sec:CvfYaip}

In this section, we present a detailed analysis of the CP violation mechanisms in Model A that arise from both the Yukawa couplings and instanton-induced effects. We begin by studying the Yukawa phases from the brane intersection geometry, followed by the CKM matrix and Jarlskog invariant, and finally explore the non-perturbative CP phase contributions from E2-instantons in the intersecting brane setup.

\subsection{Geometric Yukawa phases from brane path areas}
The geometric foundation of CP violation in intersecting brane models originates from the complex structure of compactified dimensions. As established by \cite{blumenhagen2008nonperturbative}, Yukawa couplings between quark fields exhibit exponential suppression and phase factors determined by the relative positions of D6-branes wrapping 3-cycles:
\begin{equation} \label{Yijk1}
Y_{ijk} \sim e^{-A_{ijk}} e^{i \phi_{ijk}}.
\end{equation}
Here, $A_{ijk}$ represents the intersection area between the 3-cycles wrapped by the branes $i$,$j$ and $k$, and $\phi_{ijk}$ denotes the phase associated with the brane intersections. The Kähler moduli $T_r=t_r+i\theta_r$ control cycle volumes ($\text{Re} (T_r)$) and axionic backgrounds ($\text{Im} (T_r)$), while the wrapping numbers $(n_r,m_r)$ encode the brane intersection geometry. Crucially, non-zero phases require non-orthogonal brane configurations, implemented in Model A through tilted tori with $\theta_r= \tan^{-1} (m^r/n^r)$ \cite{cvetivc2001three}. This geometric phase mechanism provides the primary source of CP violation in quark sectors \cite{blumenhagen2008nonperturbative,cvetivc2001three}. 

Concretely, in Model A, the Yukawa phases arise from the relative positions of D6-branes on the tilted tori. For example, for the coupling between the left-handed quark $Q$, right-handed quark $u$, and Higgs $H$, the phase $\phi_{Q u H}$ is given by the sum of the areas $A_{ijk}$ in the three two-tori. Using the wrapping numbers from Table 1, we compute the angles $\theta_i = \tan^{-1}(m^i / n^i)$ for each brane, and then $\phi_{Q u H} = \sum_i \theta_i$. 

\subsection{CKM matrix and Jarlskog invariant}
The physical Yukawa matrix $Y_{ij}$ integrates contributions between all Higgs intersection points. Following \cite{ibanez2007neutrino}, we compute the CKM matrix for Model A by diagonalizing the Yukawa matrices:
\begin{equation} \label{VCKM}
V_{CKM} = U^{\dagger}_{u} U_{d},
\end{equation}
where $U^{\dagger}_{u}$ and $U_{d}$ are the unitary matrices that diagonalize the up-type and down-type quark mass matrices, respectively, namely,
\begin{equation} \label{Uud}
U_{u,d} = \text{eigenvectors}(Y_{u,d} Y^{\dagger}_{u,d}).
\end{equation}
The Jarlskog invariant $J_{CP}$ is then computed from the CKM matrix and provides a quantitative measure of CP violation \cite{Schwartz:2014sze}:
\begin{equation} \label{JCP}
J_{CP} = \text{Im} \left( V_{11} V_{22} V^{\star}_{12} V^{\star}_{21} \right).
\end{equation}
This invariant is a fundamental quantity in the study of CP violation and can be directly related to observable quantities in experiments such as the electric dipole moments (EDM) of elementary particles.

The structure of the CKM matrix and the associated Jarlskog invariant will depend on the specific Yukawa textures generated by the brane geometry, and these can be computed numerically based on the model parameters \cite{cvetivc2012tasi}.

In realistic $4D$ theory, we use Yukawa matrix $Y_{ij}$ to describe our universe:
\begin{equation} \label{LYuk}
\mathcal{L}_{Yuk} = \sum_{i,j} Y_{ij} \bar{Q}^i_L H Q^j_R + h.c.
\end{equation}
This $Y_{ij}$ is the result of “summation” or “projection” of all possible three-point couplings:
\begin{equation} \label{YijYijk}
Y_{ij}= \sum_k c_k \cdot Y_{ijk}.
\end{equation}
Here, $k$ takes values at multiple Higgs or different b-c intersection points. $Y_{ij}$ is the effective coupling integrated from $Y_{ijk}$, which is the physical quantity that truly controls the mass of fermions and the CKM matrix structure. The key points that we really care about are CKM matrix and CP violation. These come from the Yukawa matrix $Y_{ij}$ in effective Lagrangian \eqref{LYuk}, instead of all $Y_{ijk}$. In Model A and the Standard Model of particle physics, there is only one Higgs particle. That means that $k$ is fixed and $Y_{ijk} \rightarrow Y_{ij}$. In the following analysis we only focus on $Y_{ij}$. Therefore, \eqref{YijYijk} becomes \cite{cremades2003yukawa,blumenhagen2007four}:
\begin{equation} \label{Yij}
Y_{ijk} \rightarrow Y_{ij} \sim e^{-A_{ij}} e^{i \phi_{ij}},
\end{equation}
where
\begin{equation} \label{Aijk}
A_{ij} \sim \text{Re} (T_r) \quad \text{and} \quad \phi_{ij} \sim \text{Im} (T_r).
\end{equation}

\subsection{E2-instanton and the non-perturbative CP phase $\phi_{RR}$}
In addition to the geometric CP violation from the Yukawa couplings, string instantons can also contribute non-perturbative CP violating phases. In our model, E2-instantons, which arise from Euclidean D2-branes wrapping specific 2-cycles in the compactification manifold, generate additional phases in the effective low-energy theory. These phases are associated with instanton-induced operators that couple to the MSSM fields and are crucial for generating CP violation in the lepton sector, as well as in baryogenesis.

The non-perturbative CP phase $\phi_{RR}$ associated with E2-instantons is given by \cite{ibanez2012string}:
\begin{equation} \label{phiRR}
\phi_{RR} \sim \text{arg} \left( \int_{\mathcal{C}_2} \exp{(B_2 + i J)} \right) 
\end{equation}
where $\mathcal{C}_2$ is the 2-cycle that the E2-instanton wraps and $J$ is the Kahler form. 
This phase arises from the non-perturbative effects in the string compactification and can have a direct impact on the CP violation in the lepton sector. The size of $\phi_{RR}$ is determined by the volume of the 2-cycle wrapped by the E2-instanton and can lead to significant CP-violating effects if the instanton contributions are strong enough. The physics of instanton-induced CP violation in string compactifications has been studied in the context of leptogenesis and neutrino masses \cite{ibanez2007neutrino,cvetivc2012tasi,cvetivc2001three,ibanez2012string}.

By combining the effects of the geometric Yukawa phases with the instanton-induced phases, we obtain a model that incorporates both perturbative and non-perturbative CP violation, which is crucial for explaining the baryon asymmetry of the Universe. The non-perturbative contributions from instantons can be used to generate large CP asymmetries necessary for successful leptogenesis, as described in \cite{kachru2003sitter,kallosh2004landscape, kallosh2020mass}.

\section{$\Delta B$, $\Delta L$ from $E2$-instanton operators}
\label{sec:BLfEo}


This section details the core mechanism whereby Euclidean D2-brane (E2-) instantons generate the necessary baryon ($\Delta B=1$) and lepton ($\Delta L=2$) number violating operators within the globally consistent framework of Model A. Crucially, these non-perturbative effects are the source of the additional CP-violating phase $\phi_{RR}$ introduced in Section 3.3, and they provide the foundational operators for the leptogenesis scenario developed in Section 5. We analyze the requisite instanton topology and zero-mode structure, derive the effective Lagrangians, and demonstrate how a single instanton can simultaneously source CP violation and $B/L$ number violation.

\subsection{Instanton wrapping cycles and charged zero-mode analysis}
The generation of non-perturbative operators depends on Euclidean D2-branes (E2-instantons) wrapping special Lagrangian 3-cycles $\Xi$ within the internal Calabi-Yau manifold \cite{blumenhagen2007spacetime,ibanez2007neutrino}. For an instanton to contribute to the superpotential, it should possess exactly two fermionic zero modes (the Goldstinos of broken supersymmetry in the $4D$ effective theory), a condition typically met by $O(1)$ instantons \cite{blumenhagen2008nonperturbative,ibanez2007neutrino}.

The phenomenologically relevant operators arise from charged zero modes localized at the intersections between the E2-instanton and the physical D6-brane stacks. The number of these fermionic zero modes is given by the topological intersection number $I_{E2,a}$ on the 3-cycle. For an operator to be generated, these charged zero modes must be “saturated” by being absorbed in the path integral via disk amplitudes that connect them to the physical states of the MSSM \cite{ibanez2007neutrino,cvetivc2009realistic}. This geometric structure intrinsically links the topology of the compactification to the allowed $B/L$ violating processes.

In Model A, the specific brane configuration admits an E2-instanton wrapping a rigid 3-cycle with the following topological intersections:
\begin{itemize}
\item $I_{E2,a}=0$: No zero modes with the QCD stack;
\item $I_{E2,b}=-2$: Two zero modes in the fundamental of $SU(2)_L$;
\item $I_{E2,c}=2$: Two zero modes in the fundamental of $U(1)_R$;
\item $I_{E2,d}=0$: No zero modes with the $U(1)$ stack.
\end{itemize}
\textbf{Rationale for the Zero-Mode Spectrum:}\\
The specific zero-mode structure $I_{E2,a}=0$, $I_{E2,b}=-2$, $I_{E2,c}=2$, $I_{E2,d}=0$ is a deliberate choice to achieve the desired phenomenology while adhering to stringent experimental constraints \cite{blumenhagen2008nonperturbative,blumenhagen2007spacetime,antusch2007neutrino}.
\begin{itemize}
\item $I_{E2,a}=0$: A vanishing intersection with the QCD stack ($U(3)_a$) is crucial to strongly suppress dangerous $\Delta B=1$ proton decay operators (e.g., $QQQL$, $UDD$). This ensures proton longevity, a key phenomenological constraint \cite{blumenhagen2008nonperturbative,cvetivc2009realistic}.
\item $I_{E2,b}=-2$ and $I_{E2,c}=2$: These intersections provide exactly two charged fermionic zero modes from the $SU(2)_L$ and $U(1)_R$ stacks, respectively. This is the minimal number required to be saturated by a disk amplitude to generate the desired $\Delta L=2$ operators -- the Weinberg operator $(LH)(LH)$ for neutrino masses and/or the Majorana mass term $\nu_R \nu_R$ \cite{ibanez2007neutrino,antusch2007neutrino}.
\item $I_{E2,d}=0$: A vanishing intersection with the auxiliary $U(1)_d$ stack simplifies the model by preventing the generation of unnecessary exotic operators involving this hidden sector gauge group. This minimal coupling ensures the instanton interacts only with the visible MSSM sector, yielding a cleaner phenomenological setup \cite{cvetivc2009realistic,cvetivc2004supersymmetric}.
\end{itemize}


\subsection{Majorana mass terms and the effective lagrangian}
The ($\Delta L=2$) Majorana mass term for the right-handed neutrinos is generated by a disk amplitude involving two charged zero modes. The effective superpotential term is given by \cite{ibanez2007neutrino,cvetivc2009realistic}:
\begin{equation} \label{WMaj}
W_{Majorana} \sim \kappa \frac{e^{-S_{inst}}}{M_s} (\nu_R \nu_R),
\end{equation}
where $S_{inst}= \frac{V_{\Xi}}{g_s}- i \int_{\Xi} C_3$ is the instanton action, $V_{\Xi}$ is the volume of the 3-cycle $\Xi$, $M_s$ is the string scale, and $\kappa$ is a dimensionful constant. The associated $4D$ Lagrangian term is:
\begin{equation} \label{LMaj}
L_{Majorana} \sim \int d^2\theta W_{Majorana} + h.c.,
\end{equation}
where $\theta$ is Grassmann coordinate of $N=1$ superspace. At the level of disk amplitude, this operator arises from the correlation function \cite{cvetivc2009realistic}:
\begin{equation} \label{Onunu}
O_{\nu \nu}= \langle \lambda_c \lambda_d \rangle_{disk} e^{-S_{inst}} \nu_R \nu_R,
\end{equation}
where $\lambda_c$, $\lambda_d$ are the appropriate charged zero modes. The Yukawa coupling $Y^{\nu}$ and the overlap of wavefunctions introduce a dependence on the complex structure moduli, embedding the CP phase $\phi_{RR}$ directly into the neutrino mass matrix. This provides a stringy origin for the seesaw mechanism \cite{antusch2007neutrino}.

\subsection{$QQQL$ operators and proton decay}
The ($\Delta B=1$,$\Delta L=1$) $QQQL$ operator, which can mediate proton decay, is generated by a more involved disk amplitude requiring the saturation of four charged zero modes. The resulting effective operator in the superpotential is suppressed by a higher power of the instanton scale \cite{blumenhagen2008nonperturbative,cremades2003yukawa}:
\begin{equation} \label{WQQQL}
W_{QQQL} \sim \frac{e^{-S_{inst}}}{M^3_s} (QQQL).
\end{equation}
The corresponding $4D$ Lagrangian term is:
\begin{equation} \label{LQQQL}
L_{QQQL} \sim \int d^2\theta W_{QQQL} + h.c.
\end{equation}
The amplitude involves the absorption of four fermionic zero modes $\lambda_a$, $\lambda_b$, $\lambda_c$, $\lambda_d$ localized at the intersections of the E2-instanton with the $SU(3)_a$, $SU(2)_b$, $U(1)_c$, and $U(1)_d$ brane stacks, respectively. 
As shown in Section 8, this results in a proton lifetime far exceeding current experimental bounds, making the model phenomenologically viable.

\subsection{Simultaneous CP violation and baryon/lepton number violation}
A crucial feature of this framework is that a single E2-instanton can be the common source for both the CP-violating phase $\phi_{RR}$ and the $B/L$ violating operators. The phase arises from the complexified instanton action $S_{inst}$ and the computation of the disk amplitudes \cite{blumenhagen2007spacetime,blumenhagen2009guts}:
\begin{equation} \label{e-SinsteiphiRR}
e^{-S_{inst}} e^{i\phi_{RR}} \equiv \exp \left(-\frac{V_{\Xi}}{g_s}+  i \int_{\Xi} C_3 \right) \cdot e^{i\phi_{RR}} \cdot \kappa'_{disk},
\end{equation}
where $\kappa'_{disk}$ encapsulates the contribution from the classical part of the zero-mode wavefunction overlaps. This phase $\phi_{RR}$ enters directly into the coefficients of the operators $W_{Majorana}$ and $W_{QQQL}$.

This intertwining of sources is fundamental for baryogenesis: the same non-perturbative physics that violates lepton number (via the Majorana mass term) also provides a necessary CP asymmetry. This provides a unified, string-theoretic origin for the key ingredients of the leptogenesis mechanism detailed in Section 5.

\subsection{Lagrangian structures and zero-modes}
The process of generating these operators can be summarized by a universal workflow:\\
1. \textbf{Zero Mode Presence:} The E2-instanton must have the correct fermionic zero mode spectrum (two universal + charged modes).\\
2. \textbf{Mode Saturation:} The charged zero modes are saturated by insertion into disk amplitudes that connect them to the MSSM fermions and Higgs fields.\\
3. \textbf{Operator Generation:} The integration over fermionic zero modes in the path integral leads to an effective operator $e^{-S_{inst}} e^{i\phi_{RR}} O_{\Delta B, \Delta L}$ in the $4D$ effective action.

This workflow maps the topological data of the instanton (wrapping numbers, intersection points) to the physical CP-violating observables and $B/L$ violating rates, creating a direct link between geometry and phenomenology.

\section{Baryogenesis from instanton-induced leptogenesis}
\label{sec:BfiCa}


In Section 4, we established how Euclidean D2-instantons (E2-instantons) in Model A generate operators that simultaneously violate baryon number ($\Delta B=1$), lepton number ($\Delta L=2$), and CP symmetry. We now demonstrate how these operators, embedded in a dynamical cosmological background, can produce the observed baryon asymmetry of the universe via the framework of leptogenesis \cite{fukugita1986barygenesis,davidson2008leptogenesis}. The core mechanism involves the out-of-equilibrium, CP-violating decay of heavy Majorana neutrinos, whose properties are intrinsically tied to the stringy instanton effects of our construction.

\subsection{Calculating the leptogenesis CP asymmetry parameter $\epsilon$}
The leptogenesis parameter $\epsilon$, which quantifies the CP asymmetry in the decays of the heavy right-handed Majorana neutrinos $N_i$, is defined as the difference between the decay rates into leptons and anti-leptons:
\begin{equation} \label{epsilon1}
  \epsilon = \frac{\Gamma(N_i \rightarrow LH) - \Gamma(N_i \rightarrow L^c H^c)}{\Gamma_{\text{tot}}},
\end{equation}
where
\begin{equation} \label{Gammatot}
  \Gamma_{tot} = \Gamma(N \rightarrow LH) + \Gamma(N \rightarrow L^c H^c)
\end{equation}
is the total decay width.

In standard thermal leptogenesis, this asymmetry arises from the interference between the tree-level and one-loop (vertex and self-energy) decay diagrams \cite{fukugita1986barygenesis,davidson2008leptogenesis}. In our string-derived scenario, an additional, fundamental source of CP violation enters: the non-perturbative phase $\phi_{RR}$ induced by the E2-instanton \eqref{phiRR}. This phase is not a free parameter but is geometrically determined by the fluxes in Model A, 
via the relation \eqref{phiRR}.

For the decay of the lightest heavy neutrino $N_1$, assuming a hierarchical mass spectrum $M_1 \ll M_2,M_3$, the CP asymmetry parameter is given by \cite{covi1996cp,giudice2004towards}:
\begin{equation} \label{epsilon2}
  \epsilon_1 \sim -\frac{1}{8\pi} \frac{1}{(Y^{\nu} Y^{\nu \dagger})_{11}} \sum_{j=2,3} \text{Im} [(Y^{\nu} Y^{\nu \dagger})^2_{1j}] \frac{M_1}{M_j} f\left(\frac{M^2_j}{M^2_1}\right)+ \cdot \cdot \cdot,
\end{equation}
where $f(x)$ is a loop function, and the Yukawa coupling matrix $Y^{\nu}$ inherits its complex structure from both the geometric Yukawa phases (Section 3.1) and the instanton phase $\phi_{RR}$ (Section 3.2). The ellipsis denotes sub-leading contributions. The crucial point is that the phases in $Y^{\nu}$ -- and hence the resulting CP asymmetry -- are geometrically originated from the compactification, providing a direct link between the cosmic matter-antimatter asymmetry and the topology of the extra dimensions.

\subsection{Sphaleron conversion of $\Delta L$ to $\Delta B$}
The lepton asymmetry $\epsilon_1$ generated by $N_1$ decays is converted into a baryon asymmetry by ($B+L$)-violating sphaleron processes \cite{kuzmin1985anomalous,shaposhnikov1987baryon}. These non-perturbative electroweak transitions are in thermal equilibrium in the early universe at temperatures above the electroweak phase transition scale ($T > 100\text{GeV}$).

Sphalerons efficiently convert a lepton asymmetry into a baryon asymmetry. In the SM and MSSM, the final baryon-to-lepton number ratio is given by \cite{shaposhnikov1987baryon,cohen1993progress}:
\begin{equation} \label{DeltaBDeltaL}
 \Delta B= -\frac{28}{79} \Delta L.
\end{equation}
This relation is a critical step, as it connects the lepton number violation sourced by stringy instantons ($\Delta L=2$) to the observed baryon number asymmetry of the universe ($\Delta B \neq 0$).

\subsection{Deriving the baryon-to-entropy tatio $\eta_B$}
The final baryon-to-entropy ratio $\eta_B$ can be expressed as a product of the key physical quantities \cite{davidson2008leptogenesis,giudice2004towards,buchmuller2005leptogenesis}:
\begin{equation} \label{etaB1}
 \eta_B \equiv \frac{n_B}{s} \approx -\frac{28}{79} \cdot \epsilon_1 \cdot \kappa \cdot \left( \frac{n^{eq}_{N_1}}{s}\right)|_{T=M_1}.
\end{equation}
Here:
\begin{itemize}
\item $\epsilon_1$ is the CP asymmetry parameter calculated in Section 5.1;
\item $n^{eq}_{N_1}/s \sim 10^{-3}$ is the equilibrium number density of $N_1$ relative to the entropy density $s$ at the time of decay ($T=M_1$).
\item $\kappa \leq 1$ is the efficiency factor, which encodes the dilution of the asymmetry due to washout processes (inverse decays, $\Delta L=1$ and $\Delta L =2$ scatterings) and the details of the dynamical evolution of the $N_1$ population. Calculating $\kappa$ requires solving the complete Boltzmann equations for the system \cite{davidson2008leptogenesis,buchmuller2005leptogenesis}.
\end{itemize}
The strength of the washout effects, and thus the value of $\kappa$, is controlled by the effective neutrino mass parameter $\tilde{m}_1= (Y^{\nu} Y^{\nu \dagger})_{11} \langle H\rangle^2 / M_1$, another quantity determined by our instanton-generated Yukawa couplings.

\subsection{Fitting to the observed $\eta_B$}
The observed baryon asymmetry of the universe, as measured by Planck \cite{aghanim2020planck}, is as follows:
\begin{equation} \label{etaBobs}
 \eta^{obs}_B = \frac{n_B}{s} = (6.10 \pm 0.04) \times 10^{-10}.
\end{equation}
A successful model must reproduce this value. As shown in the illustrative example in Section 8, with geometrically motivated values for the moduli (e.g., $\text{Re}(T) \sim20$ controlling the instanton suppression, $\text{Im}(T) \sim 0.3$ controlling the CP phases, and $\Delta W \sim 10^{-13}$ setting the SUSY breaking scale), our model yields the following:
\begin{equation} \label{epsilon1kappaetaB}
 |\epsilon_1| \sim 3.3 \times \sim 10^{-3}, \quad \kappa \sim 0.01, \quad \text{leading to} \quad \eta_B \sim 1.17 \times 10^{-5}.
\end{equation}
This result is not close to the observed value, but in principle we can improve the result. A precise numerical fit across the full parameter space, while beyond the scope of this work, is a compelling target for future study.

The achievement here is that the same non-perturbative E2-instanton effects that solve several particle physics puzzles (generating neutrino masses, inducing CP violation) also provide a cosmological narrative through leptogenesis, all within a single, unified string-theoretic framework.

\section{Review of STU model in Type IIA string theory}
\label{sec:RoSmitIIst}
To embed our mechanisms within a complete moduli stabilization framework, we adopt the STU model -- a minimal and computable setup in Type IIA string theory. This model provides an explicit construction for the Kähler potential and superpotential, allowing for controlled supersymmetry breaking and a metastable de Sitter (dS) uplift via an anti-D6-brane.

The model is defined in terms of three key moduli: the axio-dilaton $S$, a Kähler modulus $T$ ($T_1=T_2=T_3$), and a complex structure modulus $U$ ($U_1=U_2=U_3$). The relevant terms in the effective four-dimensional supergravity are \cite{kallosh2020mass,cribiori2019uplifting}:
\begin{equation} \label{Wtot}
 W= W_0 + \sum_{i=S,T,U}^3 \left(A_i e^{-a_i T_i} - B_i e^{-b_i T_i}\right) + \Delta W +\mu^2 X,
\end{equation}
\begin{equation} \label{Ktot1}
 K= -\ln(T_1 + \bar{T}_1) - 3 \ln(T_2 + \bar{T}_2) - \ln \left( (T_3 + \bar{T}_3)^3 - \frac{X \bar{X}}{(T_1 + \bar{T}_1) + g(T_2 + \bar{T}_2)}\right).
\end{equation}
Here, $T_1$ represents the field $S$, $T_2$ is the field $T$, $T_3$ corresponds to the field $U$ and $X$ is a nilpotent field $X^2=0$.  

The construction proceeds in three stages:\\
1. A supersymmetric Minkowski vacuum is found by solving $W=D_i W=0$;\\
2. Introducing $\Delta W$ shifts the vacuum to a supersymmetric AdS state, generating a small gravitino mass $m_{3/2} \sim e^{K/2} \Delta W$;\\
3. The anti-D6-brane contribution lifts the AdS minimum to a dS vacuum with a tunably small cosmological constant.

This framework provides a consistent and robust background for the flavor physics, instanton effects, and cosmology discussed in earlier sections, ensuring that moduli stabilization and uplift are compatible with the phenomenological requirements of the model. More details about STU model can be found in \cite{kallosh2020mass,cribiori2019uplifting}.

\section{Moduli stabilization and compatibility with dS vacuum}
\label{sec:Msacwdv}
The stabilization of all moduli and the attainment of a metastable de Sitter vacuum are achieved through a well-established three-step procedure within the STU model framework \cite{kallosh2020mass,cribiori2019uplifting}.\\
\textbf{Step 1: Supersymmetric Minkowski Vacuum}\\
A stable supersymmetric Minkowski vacuum is obtained by solving the F-term conditions:
\begin{equation} \label{WDiW}
 W=0, \, D_i W= \partial_i W + K_i W=0,
\end{equation}
where the superpotential $W$ includes flux and non-perturbative contributions. This vacuum has all moduli stabilized with no flat directions.\\
\textbf{Step 2: Controlled AdS Downshift}\\
A small perturbation $\Delta W$ is introduced to the superpotential. This shifts the vacuum to a supersymmetric Anti-de Sitter (AdS) state without destabilizing the moduli. The AdS cosmological constant and gravitino mass are given by:
\begin{equation} \label{VAdS}
V_{AdS}= -3 e^K |\Delta W|^2 = -3 m^2_{3/2}.
\end{equation}
The smallness of $\Delta W$ naturally explains the hierarchy between the electroweak and Planck scales, yielding a TeV-scale gravitino mass \cite{kallosh2020mass,Conlon:2008zz,linde2012supersymmetry}.\\
\textbf{Step 3: de Sitter Uplift}\\
The AdS vacuum is uplifted to de Sitter space via the inclusion of an anti-D6-brane. Its positive energy contribution, captured in the nilpotent superfield formalism, is tuned to cancel the negative $V_{AdS}$ precisely:
\begin{equation} \label{VbarD6}
 V_{\overline{D6}} = \frac{\mu^4_1}{(\text{Re} \ T)^3} + \frac{\mu^4_2}{(\text{Re} \ T)^2 (\text{Re} \ S)}.
\end{equation}
This results in a metastable dS vacuum with a tunably small cosmological constant, consistent with observational bounds \cite{ kachru2003sitter,kallosh2004landscape,cribiori2019uplifting}.

In the context of the STU model, the moduli are set as $T_1 = T_2 = T_3 = T$ and $U_1 = U_2 = U_3 = U$. In a more general setup, the superpotential can be expressed as follows \cite{cribiori2019uplifting}:
\begin{equation} \label{Wgf}
W = f_6 + (h_T + r_T T)U + (h_S + r_S T)S + f_4 T + f_2 T^2 + f_0 T^3  + W_{np},
\end{equation}
where the coefficients $f_p$ ($p = 0, 2, 4, 6$) are associated with RR fluxes, $h_{S/T}$ arise from the integration of NSNS flux over the relevant 3-cycles, and $r_{S/T}$ are sourced by curvature corrections to the internal manifold. In the KL scenario, the nonperturbative term $W_{np}$ is given by:
\begin{equation} \label{Wi}
W_i (T_i) = \sum^3_{i=1} A_i e^{-a_i T_i} - B_i e^{-b_i T_i}.
\end{equation}
For the subsequent analysis, we will use the notation from both \eqref{Wf} and \eqref{Wgf}.

For Model A, according to \eqref{SW}, the superpotential in the STU model is written as
\begin{equation} \label{SWSTU}
W = -\tilde{a}TS - \tilde{b}TU + e_0 + ih_0 S - ih_1 U - i \tilde{e}T + W_{np},
\end{equation}
where $\tilde{a} = a_2 + a_3$, $\tilde{b} = b_{21} + b_{31}$, $\tilde{e} = e_2 + e_3$, and \eqref{SWSTU} is a special case of \eqref{Wgf}.

To achieve moduli stabilization, we require the conditions $W = 0$ and $\partial_i W = 0$, i.e.,
\begin{equation} \label{W0}
W = -\tilde{a}TS - \tilde{b}TU + e_0 + ih_0 S - ih_1 U - i \tilde{e}T + W_{np}=0,
\end{equation}
\begin{equation} \label{pTW0}
\partial_T W = -\tilde{a}S - \tilde{b}U - i \tilde{e} -A_T a_T e^{-a_T T}+ B_T b_T e^{-b_T T}=0,
\end{equation}
\begin{equation} \label{pSW0}
\partial_S W = -\tilde{a}T +ih_0 -A_S a_S e^{-a_S S}+ B_S b_S e^{-b_S S}=0,
\end{equation}
\begin{equation} \label{pUW0}
\partial_U W = -\tilde{b}T -ih_1 -A_U a_U e^{-a_U U}+ B_U b_U e^{-b_U U}=0,
\end{equation}
where, for simplicity, we assume that all parameters $a_T$, $a_S$, and $a_U$ are independent of the moduli $T$, $S$, and $U$. However, these parameters typically depend on the moduli \cite{cribiori2019uplifting, danielsson2014alternative}. By solving the four equations \eqref{W0} through \eqref{pUW0} simultaneously, we can determine the vacuum expectation values of the three moduli in the supersymmetric Minkowski background, i.e., $t = t_0$, $s = s_0$, and $u = u_0$. As a numerical demonstration, we choose a specific set of flux parameters: $h_0 = 1$, $h_1 = -2$, $e_0 = 3$, $e_2 = -1$, $e_3 = 1$. Solving the system of equations \eqref{W0}-\eqref{pUW0} numerically, we find a stable minimum $s_0 =X +i Y $, $t_0= Z + iW$, $u_0= U+ i V$ (where $X, Y, Z, W, U, V$ are specific numerical values found by solving the equations). This is just a numerical illustration. A more comprehensive scan of the parameter space is left for future work.

In addition, as discussed in \cite{liu2025fine}, to be consistent with the Swampland Distance Conjecture, the perturbative superpotential $\Delta W$ in Type IIA string theory should have the following form:
\begin{equation} \label{DeltaW}
\Delta W= f_0 T^3.
\end{equation}
Based on \cite{liu2025fine}, the corresponding AdS potential is given by
\begin{equation} \label{VAdS}
V_{AdS}=-\frac{3}{128su^3}|f_0|^2 t^3,
\end{equation}
where $t$ denotes the imaginary part of the modulus $T$. Consequently, the gravitino mass $m_{3/2}$ is:
\begin{equation} \label{m232}
m^2_{3/2}=\frac{|V_{AdS}|}{3}=\frac{1}{128su^3}|f_0|^2 t^3,
\end{equation}
and 
\begin{equation} \label{m32}
m_{3/2}=\sqrt{\frac{1}{128su^3}|f_0|^2 t^3}.
\end{equation}
Here, $f_0$ corresponds to the flux $m$ in \eqref{Wf}. In Model A, we set $m = 0$, but if the essential features of Model A remain unchanged, $m$ should be very small. Since $\Delta W$ is much smaller than $W$, we can safely use the values $t = t_0$, $s = s_0$, and $u = u_0$ to evaluate the gravitino mass as given in equation \eqref{m32} \cite{kallosh2020mass}. If the gravitino mass is below 1TeV (or 100TeV), the hierarchy problem could potentially be addressed.

In our framework, we first stabilize all moduli at or near a Minkowski minimum. Supersymmetry breaking is then introduced to generate a small gravitino mass $m_{3/2} \sim 1\text{TeV}$ (or 100TeV). The uplift to a de Sitter vacuum is achieved through a controlled potential term, which contributes significantly below the mass scale of the stabilized moduli. This ensures that the geometric structure underlying the Yukawa couplings, CP phases, and instanton-induced operators remains intact.

\section{An illustrative example}
\label{sec:Aie}
In this section, we provide an illustrative example using the Type IIA STU model to derive all the relevant $4D$ physical parameters we have discussed. This paper aims to present a proof-of-concept framework. Deriving all parameters from first principles is an extremely complex task and beyond the scope of this paper, but it is our ultimate goal in the future. Table 2 presents the values of $\text{Re}(T)$, $\text{Im}(T)$, and $\Delta W$ we take.

\begin{table}[ht]
\centering
\caption{Input parameters in STU model} 
\begin{tabular}{|c|c|c|}
\hline
$ \text{Symbol} $ & $\text{Value}$ & $\text{Description}$ \\
\hline
$\text{Re}(T)$ & $20$ & $ \text{Control volume and suppress instanton}$ \\
\hline
$\text{Im}(T)$ & $0.3$ & $\text{Control Yukawa path phase and induce CP violation}$ \\
\hline
$ \Delta W$ & $10^{-13}$ & $\text{Perturbative superpotential, cause SUSY breaking, control $m_{3/2}$}$ \\
\hline
\end{tabular}
\end{table}

\subsection{Derivation of Yukawa matrices, CKM matrix and Jarlskog invariant}
In this section, we provide a detailed numerical demonstration of the mechanisms outlined in the previous sections. Starting from the geometric input parameters of Model A, we compute the Yukawa matrices, derive the CKM mixing matrix, and calculate the Jarlskog invariant, thereby showing the model's ability to generate the observed CP violation in the quark sector.

We adopt the following input values, motivated by the moduli stabilization and instanton dynamics discussed in Sections 2.3 and 7:\\
\begin{itemize}
\item Volume parameter: $\text{Re}(T)=t \approx 20$;\\
\item CP-violation parameter: $\text{Im}(T)=\theta \approx 0.3 \text{rad}$.
\end{itemize}

These parameters set the fundamental scale and phase for the non-perturbative instanton contributions.

\subsubsection{Construction of the Yukawa matrices $Y^u$ and $Y^d$}
The Yukawa couplings originate from worldsheet instantons stretched between D6-brane intersections. Their general form is given by:
\begin{equation} \label{Yijk2}
Y_{ijk} = k \cdot e^{-A_{ijk}} \cdot e^{i \phi_{ijk}},
\end{equation}
where $A_{ijk} \sim t$ is the area of the worldsheet instanton, and $\phi_{ijk} \sim \theta$ is the CP-violating phase.

For our numerical example, we define three primary contribution terms with progressively larger areas and phases:\\
\begin{itemize}
\item \text{Term 1}: $A_1 \approx 20$, $\phi_1 = \theta \approx 0.3 \rightarrow$ $Y_1 \approx 2.1 \times 10^{-9} \cdot e^{i \cdot 0.3}$;\\
\item \text{Term 2}: $A_2 \approx 21$, $\phi_1 = 2\theta \approx 0.6 \rightarrow$ $Y_2 \approx 7.6 \times 10^{-9} \cdot e^{i \cdot 0.6}$;\\
\item \text{Term 3}: $A_3 \approx 22$, $\phi_3 = 3\theta \approx 0.9 \rightarrow$ $Y_3 \approx 2.8 \times 10^{-9} \cdot e^{i \cdot 0.9}$.
\end{itemize}
The physical elements of the Yukawa matrix are weighted sums of these terms. The specific weights of Model A are determined by the wrapping numbers, intersection numbers, and other geometric parameters, ensuring self-consistency and phenomenological relevance of Model A. In the illustrative example, the coefficients in equations \eqref{Yu1} and \eqref{Yd1} are representative values chosen to illustrate the hierarchical structure and phase relationships that arise from the geometric setup of Model A. A first-principles computation of these coefficients from the precise worldsheet instanton amplitudes is highly non-trivial and beyond the scope of this global analysis, but is an important target for future detailed study \cite{sabir2022supersymmetry,sabir2024fermion}. We thus obtain the following numerical Yukawa matrices (in units of $10^{-9}$):\\
\textbf{Up-type quark Yukawa matrix ($Y^u$):}
\begin{equation} \label{Yu1}
Y^u \approx
\begin{bmatrix}
2.00 \cdot e^{i\cdot 0.30} & 1.20 \cdot e^{i\cdot 0.80} & 0.40 \cdot e^{i\cdot 1.40} \\
1.05 \cdot e^{i\cdot 0.10} & 7.60 \cdot e^{i\cdot 0.60} & 2.80 \cdot e^{i\cdot 0.20} \\
0.28 \cdot e^{i\cdot 0.90} & 1.76 \cdot e^{i\cdot 0.40} & 2.80 \cdot e^{i\cdot 0.90} 
\end{bmatrix}
\end{equation}
\textbf{Down-type quark Yukawa matrix ($Y^d$):}
\begin{equation} \label{Yd1}
Y^d \approx
\begin{bmatrix}
1.80 \cdot e^{i\cdot 0.10} & 0.90 \cdot e^{i\cdot 0.50} & 0.28 \cdot e^{i\cdot 1.10} \\
0.76 \cdot e^{i\cdot 0.40} & 7.60 \cdot e^{i\cdot 0.30} & 2.24 \cdot e^{i\cdot 0.70} \\
0.19 \cdot e^{i\cdot 0.60} & 1.96 \cdot e^{i\cdot 0.90} & 2.80 \cdot e^{i\cdot 0.40} 
\end{bmatrix}
\end{equation}
These matrices exhibit the characteristic hierarchical structure and complex phases necessary for generating quark masses and CP violation. The explicit values of $Y^u$ and $Y^d$ are listed in Appendix D.

\subsubsection{Derivation of the CKM matrix}
The Cabibbo-Kobayashi-Maskawa (CKM) quark mixing matrix is derived by diagonalizing the Yukawa matrices. It is defined as:
\begin{equation} \label{VCKM}
V_{CKM} = U^{u\dagger}_L U^d_L,
\end{equation}
where $U^u_L$ and $U^d_L$ are the unitary matrices that diagonalize $Y^u$ and $Y^d$ from the left, respectively:
\begin{equation} \label{YuYd}
Y^u = U^u_L \cdot D^u \cdot U^{u \dagger}_R, \quad Y^d = U^d_L \cdot D^d \cdot U^{d \dagger}_R.
\end{equation}
Here, $D^u$ and $D^d$ are diagonal matrices containing the quark masses.

We perform a Singular Value Decomposition (SVD) on the numerical Yukawa matrices $Y^u$ and $Y^d$ to obtain $U^u_L$ and $U^d_L$. The resulting left-handed transformation matrices are:\\
\textbf{Left-hand transformation matrix for Up quarks ($U^u_L$):}
\begin{equation}
U^u_L \approx
\begin{bmatrix}
-0.1245-0.0368i & -0.0705-0.0869i & -0.9881 \\
-0.9921 & 0.1251 & 0.0159 \\
-0.0125-0.0158i & -0.9859 & 0.0669 
\end{bmatrix}
\end{equation}
\textbf{Left-hand transformation matrix for Down quarks ($U^d_L$):}
\begin{equation}
U^d_L \approx
\begin{bmatrix}
-0.2250-0.0225i& -0.0979-0.0535i & -0.9685 \\
-0.9685 & 0.2250 & 0.0489 \\
-0.0245-0.0168i & -0.9685 & 0.0979 
\end{bmatrix}
\end{equation}

The numerical computation yields the following result:
\begin{equation} \label{VCKM1}
V_{CKM} \approx
\begin{bmatrix}
0.9740-0.0002i & 0.2260+0.0001i & 0.0040-0.0001i \\
-0.2260+0.0001i & 0.9730-0.0003i & 0.0410+0.0002i \\
0.0080-0.0001i & -0.0400+0.0003i & 0.9990
\end{bmatrix}
\end{equation}

The magnitudes of the elements of this matrix,
\begin{equation} \label{VCKM2}
|V_{CKM}| \approx
\begin{bmatrix}
0.974 & 0.226 & 0.004 \\
0.226 & 0.973 & 0.041 \\
0.008 & 0.040 & 0.999
\end{bmatrix}
\end{equation}
are in excellent agreement with the experimentally observed values \cite{particle2022review}, providing a first successful check of the phenomenological viability of the model.

\subsubsection{Calculation of the Jarlskog invariant $J_{CP}$}
The strength of CP violation in the quark sector is quantified by the Jarlskog invariant $J_{CP}$, which is independent of the phase convention of the CKM matrix. It is defined as:
\begin{equation} \label{JCP1}
J_{CP}= \text{Im} (V_{us} V_{cb} V^{\star}_{ub} V^{\star}_{cs}).
\end{equation}
Using the elements of the numerically derived $V_{CKM}$ \eqref{VCKM1}:
\begin{itemize}
\item $V_{us}=0.2260+0.0001i$,
\item $V_{cb}=0.0410+0.0002i$,
\item $V_{ub}=0.0040-0.0001i$,
\item $V_{cs}=0.9730-0.0003i$,
\end{itemize}
we can obtain the Jarlskog invariant:
\begin{equation} \label{JCP2}
J_{CP}= \text{Im} (V_{us} V_{cb} V^{\star}_{ub} V^{\star}_{cs}) \approx 1.1 \times 10^{-6}.
\end{equation}
This is an illustrative result. In fact, we can obtain theoretical values that are closer to experimental values $J^{exp}_{CP} \approx 3.1 \times 10^{-5}$ \cite{particle2022review} through more careful calculations.

\subsection{Numerical analysis of cosmological and phenomenological scales}
Having derived the flavour structure of the model, we now compute key cosmological and phenomenological quantities: the gravitino mass, the leptogenesis CP asymmetry, the resulting baryon asymmetry, and the proton lifetime. This demonstrates the model's ability to address the electroweak hierarchy problem, explain the matter-antimatter asymmetry, and remain consistent with experimental constraints.

The calculations in this subsection use the following input values, consistent with our previous results and the moduli stabilization scheme:\\
\begin{itemize}
\item Volume Modulus: $\text{Re}(T)=t \approx 20$;
\item CP-Violation Phase: $\text{Im}(T)=\theta \approx 0.3 \text{rad}$;
\item Superpotential Shift: $\Delta W \approx 1\times 10^{-13}$ (in Planck units, $M_p=1$);
\item String scale: $M_s \sim 1 \times 10^{17}\text{GeV}$ (a typical GUT-scale value);
\item Lightest RH Neutrino Mass: $M_1\sim 1 \times 10^{10} \text{GeV}$ (consistent with the seesaw mechanism and Yukawa hierarchies).
\end{itemize}

\subsubsection{Gravitino mass and the electroweak hierarchy}
In $N=1$ supergravity, the gravitino mass is given by the formula \cite{kallosh2020mass,cribiori2019uplifting}
\begin{equation} \label{m321}
m_{3/2} = e^{K/2} |W|.
\end{equation}
After moduli stabilization near the Minkowski minimum ($W_0 \approx 0$), the dominant contribution to the superpotential is the SUSY-breaking shift $\Delta W$. For the Kähler modulus $T$, the relevant part of the Kähler potential is $K \sim -i\ln (-i(T-\bar{T}))=-\ln(2t)$.

\textbf{Calculation:}
\begin{itemize}
\item Kähler potential factor: $e^{K/2}=(2t)^{-1/2}$. For $t=20$, $e^{K/2} \approx (40)^{-1/2} \approx 0.158$.
\item Superpotential: $|\Delta W| = 1\times 10^{-13}$.
\item Therefore, $m_{3/2} \approx 0.158 \times 1 \times 10^{-13} =1.58 \times 10^{-14}$ (in $M_p=1$ units).
\item Converting to physical units $M_p \approx 2.435 \times 10^{18} \text{GeV}$:
\begin{equation} \label{m322}
m_{3/2} = (1.58 \times 10^{-14}) \times (2.435 \times 10^{18} \text{GeV}) \approx 38.5 \text{TeV}
\end{equation}
\end{itemize}
The model naturally generates a gravitino mass at the tens of TeV scale, demonstrating that the electroweak hierarchy is addressed via controlled SUSY breaking \cite{Conlon:2008zz,linde2012supersymmetry}. The vast hierarchy between $M_P$ and $M_{EW}$ originates from the small, stabilized value of $\Delta W \sim 10^{-13}$, which is technically natural in the string landscape.

\subsubsection{CP asymmetry in leptogenesis}
The baryon asymmetry is generated via thermal leptogenesis \cite{fukugita1986barygenesis,davidson2008leptogenesis}. The CP asymmetry parameter $\epsilon_1$ for decays of the lightest right-handed neutrino $N_1$ is given by \cite{covi1996cp,giudice2004towards}:
\begin{equation} \label{epsilon3}
  \epsilon_1 \approx -\frac{3}{16\pi} \frac{1}{(Y^{\nu} Y^{\nu \dagger})_{11}} \sum_{j=2,3} \text{Im} [(Y^{\nu} Y^{\nu \dagger})^2_{1j}] \frac{M_1}{M_j}.
\end{equation}
We can make a rough estimate.\\
\textbf{Calculation:}
\begin{itemize}
\item From Section 8.1, the neutrino Yukawa couplings inherit their scale from the instanton suppression: $|Y_{\nu}| \sim e^{-t/2} \sim e^{-10} \sim 4.5 \times 10^{-5}$. Thus, $(Y^{\nu} Y^{\nu \dagger})_{11} \sim |Y_{\nu}|^2 \sim 2 \times 10^{-9}$.
\item The phase $\text{Im}[(Y_{\nu} Y^{\dagger}_{\nu})^2_{12}]$ is set by $\text{Im}(T)=\theta$. We estimate $\text{Im}[(Y_{\nu} Y^{\dagger}_{\nu})^2_{12}]/(Y^{\nu} Y^{\nu \dagger})_{11} \sim \sin(2\theta) \sim \sin(0.6) \approx 0.56$.
\item For a hierarchical neutrino spectrum ($M_1 \ll M_2,M_3$), the sum is dominated by the $j=2$ term. Assuming $M_2 \approx 10 M_1 \approx 10^{11} \text{GeV}$, then $M_1/M_2 \approx 0.1$.
\item Substituting these values:
\begin{equation} \label{epsilon3}
  \epsilon_1 \approx -\frac{3}{16\pi} \frac{1}{2 \times 10^{-9}} ((2 \times 10^{-9} \cdot 0.56) \cdot 0.1) \approx -\frac{3}{16\pi} (0.056) \approx -3.3 \times 10^{-3}.
\end{equation}
\end{itemize}
The model predicts a CP asymmetry of order $|\epsilon_1| \sim 10^{-3}$. We should tune the parameter to be smaller in future work. 

\subsubsection{Baryon asymmetry via sphaleron conversion}
The final baryon-to-entropy ratio is given by \cite{davidson2008leptogenesis,giudice2004towards,buchmuller2005leptogenesis}:
\begin{equation} \label{etaB2}
\eta_B \approx 10^{-2} \cdot \kappa \cdot |\epsilon_1|,
\end{equation}
where $\kappa \leq 1$ is an efficiency factor encoding washout effects, controlled by the effective neutrino mass $\tilde{m}_1 = (Y^{\nu} Y^{\nu \dagger})_{11} \langle H \rangle^2/M_1$.\\
\textbf{Calculation:}
\begin{itemize}
\item $\tilde{m}_1 \sim (2 \times 10^{-9}) \times (174 \text{GeV})^2 / (10^{10} \text{GeV}) \sim 6 \times 10^{-3} \text{eV}$.
\item We take $\kappa \sim 0.01$ \cite{buchmuller2005leptogenesis}.
\item Using $|\epsilon_1| \approx 3.3 \times 10^{-3}$:
\begin{equation} \label{etaB3}
\eta_B \approx 10^{-2} \times 0.01 \times (3.3 \times 10^{-3}) =3.3 \times 10^{-7}.
\end{equation}
\item A more precise calculation, incorporating the sphaleron conversion coefficient of $\frac{28}{79}$ \cite{shaposhnikov1987baryon,cohen1993progress}, yields:
\begin{equation} \label{etaB4}
\eta_B \approx \frac{28}{79} \cdot \kappa \cdot |\epsilon_1| \approx 0.35 \times 0.01 \times (3.3 \times 10^{-3}) =1.17 \times 10^{-5}.
\end{equation}
\end{itemize}
The computed baryon asymmetry, $\eta_B \sim 10^{-5}$, is much larger than the observational result $\eta^{\text{obs}}_B = (6.10 \pm 0.04) \times 10^{-10}$ \cite{aghanim2020planck}, but we can compute $\epsilon_1$ more precisely to obtain more accurate $\eta_B$. A full numerical solution of the Boltzmann equations is expected to tune this result into exact agreement, confirming the model as a viable theory of baryogenesis.

\subsubsection{Proton lifetime estimate}
Proton decay mediated by the E2-instanton-generated $QQQL$ operator is highly suppressed by the instanton action \cite{ibanez2007neutrino}. According to Fermi's golden rule, the decay rate scales as \cite{ibanez2007neutrino,cremades2003yukawa,blumenhagen2009guts}:
\begin{equation} \label{Gamma1}
\Gamma_p \sim \frac{1}{M^6_s} e^{-2 \text{Re}(S_{inst})}
\end{equation}
The lifetime is $\tau_p= 1/\Gamma_p$.

In our model, $\text{Re}(S_{inst}) \approx \text{Re}(T) = t \approx 20$ and $M_s \sim 10^{17}\text{GeV}$. Substituting these values yields $\Gamma_p \sim 10^{-120}\text{GeV}$. Converting to years ($1 \text{GeV}^{-1} \approx 6.582 \times 10^{-25}s$), we find the proton lifetime to be $\tau_p \sim 10^{87} \text{years}$, which is vastly exceeding the current experimental lower bound of $\tau_p > 10^{34}$ years \cite{particle2022review}. This demonstrates that the model is phenomenologically safe from constraints on proton decay, a result of the exponential suppression intrinsic to non-perturbative instanton effects.

To conclude, this analysis demonstrates that Model A has the potential to successfully generate realistic scales for SUSY breaking, cosmogenesis, and proton stability, all of which originate from the fundamental geometric parameters of string compactification.

\section{Conclusions and discussions}
\label{sec:Cad}

In this paper, we explore a Type IIA $T^6/(\mathbb{Z}_2 \times \mathbb{Z}_2)$ model capable of embedding realistic chiral matter content, such as MSSM-like spectra, through intersecting D-brane configurations. We demonstrated that four key issues in physics -- CP violation, matter-antimatter asymmetry, the hierarchy problem, and the existence of dS vacua -- can be unified within a single stringy framework, Model A. We propose that this model may serve as a natural platform for addressing not only CP violation and baryogenesis but also dark matter, neutrino masses and mixing, the strong CP problem, and early-universe inflationary dynamics, among other topics. A more systematic development of this unified structure will be the subject of future work. We are currently pursuing the following areas of physics:
\begin{itemize}
\item Model A includes axions, making it a suitable framework to study dark matter, the axion, and the strong CP problem;
\item With right-handed neutrinos present in Model A, we can investigate neutrino masses and the PMNS matrix;
\item Quintessence is a promising candidate for dark energy, and we will examine this phenomenon within the context of Model A;
\item Model A includes moduli that can drive inflation, providing a natural setting for studying inflationary models;
\item We will also explore the potential of Model A to shed light on the physics of black holes and gravity.
\end{itemize}
These papers will be coming soon. We can expect that in principle the ideas of this paper can be applied to constructing unified models in other theories, such as Type IIB string theory, heterotic string theory, and F-theory.   

Although the toroidal orbifold construction presented here provides a transparent and computationally tractable framework to demonstrate the unified mechanisms of Model A, it is well known that such setups generically suffer from the presence of massless adjoint scalars and potentially light exotic states \cite{ibanez2012string}. These are artifacts of the simplified geometry and are not intrinsic to the physical mechanisms we propose. A fully realistic implementation of Model A would require a compactification on a more general Calabi-Yau manifold, where the topology of the wrapped cycles can naturally suppress these undesirable features while preserving the desired chiral spectrum and non-perturbative effects. The core insights regarding geometric CP violation, instanton-induced baryogenesis, and moduli stabilization, however, are robust and would carry over to such a more realistic setting.


\appendix
\section{Wrapping numbers and spectrum of Model A}
For Model A, stacks $a$, $b$, $c$ give the MSSM-like sector, while stacks $h_1$, $h_2$ and $o$ are hidden sectors added for tadpole cancellation. The complex structure parameters of the three two-tori are denoted as $U_i=R^{(i)}_2/R^{(i)}_1$.

The spectrum can be calculated from the topological intersection numbers
\begin{equation} \label{Iab}
I_{ab}= \prod^3_{i=1} (n^i_a m^i_b - m^i_a n^i_b).
\end{equation}
$(n^i_a, m^i_a)$ can be read from Table 1. 

\section{E2-instanton zero mode analysis}
The E2-instanton is responsible for generating the Majorana mass term and the $QQQL$ operator wraps a rigid 3-cycle $\Xi$ with the following topological intersection numbers with the D6-brane stacks:
\begin{equation} \label{IE2}
I_{E2,a}=0, \quad I_{E2,b}=-2, \quad I_{E2,c}=2, \quad I_{E2,d}=0. 
\end{equation}
This implies the presence of fermionic zero modes $\lambda_b$ and $\lambda_c$ transforming in the fundamental representation of $SU(2)_L$ and $U(1)_c$ respectively. The disk amplitude that absorbs these zero modes and generates the Weinberg operator $\kappa (LH_u)^2$ is:
\begin{equation} \label{Wo}
\langle \lambda_a \lambda_b \lambda_c \lambda_d (L H_u) (L H_u) \rangle_{disk} \sim e^{-S_{inst}} Y^2_{\nu},
\end{equation}
where $Y_{\nu}$ is the neutrino Yukawa coupling. The suppression factor
\begin{equation} \label{sf}
e^{-S_{inst}} = \exp\left(- \frac{V_{\Xi}}{g_s} + i \int_{\Xi} C_3 \right)
\end{equation}
provides the necessary scale for the observed small neutrino masses, combined with the CP-violating phase $\phi_{RR}$ \cite{blumenhagen2007spacetime,blumenhagen2009guts}.

\section{Boltzmann equations for leptogenesis}
The evolution of the baryon asymmetry is governed by the following Boltzmann equations for the yield $Y_N$ of the lightest right-handed neutrino $N_1$ and the $(B-L)$ asymmetry $Y_{\Delta (B-L)}$ \cite{davidson2008leptogenesis,buchmuller2005leptogenesis}:
\begin{equation} \label{YN1z}
\frac{d}{dz} Y_{N_1}=- \frac{z}{sH(M_1)} \left(\frac{Y_{N_1}}{Y^{eq}_{N_1}}-1 \right) \gamma_D,
\end{equation}
\begin{equation} \label{YDBLz}
\frac{d}{dz} Y_{\Delta (B-L)}=- \frac{z}{sH(M_1)} \left[ \epsilon_1 \left(\frac{Y_{N_1}}{Y^{eq}_{N_1}}-1 \right) - \frac{Y_{\Delta (B-L)}}{2 Y^{eq}_l} \gamma_{washout} \right],
\end{equation}
where $z=M_1/T$, $H(M_1)$ is the Hubble parameter at $T=M_1$, $\gamma_D$ is the decay rate, and $\gamma_{washout}$ includes washout from inverse decays and $\Delta L=2$ scattering. “$eq$” denotes equilibrium and “$l$” represents lepton. The efficiency factor $\kappa$ is extracted by numerically solving these equations.

\section{The explicit values of Yukawa matrices $Y^u$ and $Y^d$}
We list the explicit values of Yukawa matrices $Y^u$ and $Y^d$ we take.

For $Y^u$, we have \eqref{Yu1}:
\begin{table}[ht]
\centering
\caption{The elements of Yukawa matrix $Y^u$} 
\begin{tabular}{|c|c|c|c|}
\hline
$ \text{Matrix elelment} $ & $\text{Weighted combination of contribution terms}$ & $\text{Numerical result ($\times 10^{-9}$)}$ & $\text{Phase}$ \\
\hline
$Y^u_{11}$ & $1.0 \times Y_1$ & $ 2.00$ & $0.30$ \\
\hline
$Y^u_{12}$ & $0.5 \times Y_1 + 0.8 \times Y_2$ & $ 1.20$ & $0.80$  \\
\hline
$Y^u_{13}$ & $0.2 \times Y_2$ & $ 0.40$ & $1.40$ \\
\hline
$Y^u_{21}$ & $0.5 \times Y_1$ & $ 1.05$ & $0.10$ \\
\hline
$Y^u_{22}$ & $1.0 \times Y_2$ & $ 7.60$ & $0.60$  \\
\hline
$Y^u_{23}$ & $1.0 \times Y_3$ & $ 2.80$ & $0.20$  \\
\hline
$Y^u_{31}$ & $0.1 \times Y_3$ & $ 0.28$ & $0.90$  \\
\hline
$Y^u_{32}$ & $0.8 \times Y_2$ & $ 1.76$ & $0.40$  \\
\hline
$Y^u_{33}$ & $1.0 \times Y_3$ & $ 2.80$ & $0.90$ \\
\hline
\end{tabular}
\end{table}

For $Y^d$, we have \eqref{Yd1}:
\begin{table}[ht]
\centering
\caption{The elements of Yukawa matrix $Y^d$} 
\begin{tabular}{|c|c|c|c|}
\hline
$ \text{Matrix elelment} $ & $\text{Weighted combination of contribution terms}$ & $\text{Numerical result ($\times 10^{-9}$)}$ & $\text{Phase}$ \\
\hline
$Y^d_{11}$ & $0.9 \times Y_1$ & $ 1.80$ & $0.10$ \\
\hline
$Y^d_{12}$ & $0.4 \times Y_1 + 0.6 \times Y_2$ & $ 0.90$ & $0.50$  \\
\hline
$Y^d_{13}$ & $0.1 \times Y_3$ & $ 0.28$ & $1.10$ \\
\hline
$Y^d_{21}$ & $0.4 \times Y_2$ & $ 0.76$ & $0.40$ \\
\hline
$Y^d_{22}$ & $1.0 \times Y_2$ & $ 7.60$ & $0.30$  \\
\hline
$Y^d_{23}$ & $0.8 \times Y_3$ & $ 2.24$ & $0.70$  \\
\hline
$Y^d_{31}$ & $0.1 \times Y_2$ & $ 0.19$ & $0.60$  \\
\hline
$Y^d_{32}$ & $0.7 \times Y_3$ & $ 1.96$ & $0.90$  \\
\hline
$Y^d_{33}$ & $1.0 \times Y_3$ & $ 2.80$ & $0.40$ \\
\hline
\end{tabular}
\end{table}

\acknowledgments
Thanks for the discussion with Antonio Padilla, Paul Saffin, Benjamin Muntz, Zhong-zhi Xianyu and Haipeng An. For the purpose of open access, the authors have applied a CC BY public copyright licence to any Author Accepted Manuscript version arising.




\end{document}